\documentclass[12pt]{article}

\AtBeginDocument{
  \addtocontents{toc}{\footnotesize}
}

\pdfoutput=1

\usepackage{amsmath,amssymb,amsfonts,amscd,mathrsfs}
\usepackage{xcolor}
\definecolor{darkblue}{rgb}{0.1,0.1,.7}
\usepackage[colorlinks, linkcolor=darkblue, citecolor=darkblue, urlcolor=darkblue, linktocpage]{hyperref} 
\usepackage[square, comma, compress,numbers]{natbib}
\usepackage[]{version}
\usepackage[]{graphicx}
\usepackage[]{latexsym}
\usepackage{geometry}
\usepackage[all,cmtip]{xy}
\usepackage[margin=10pt,font=small,labelfont=bf]{caption}
\usepackage{ifthen}
\usepackage{tikz}
\usepackage{array,setspace,mathrsfs,amsfonts,yfonts,dsfont,bbm,colonequals}

\def\be{\begin{equation}}
\def\ee{\end{equation}}
\def\bea{\begin{eqnarray}}
\def\eea{\end{eqnarray}}

\def\nn{\nonumber}
\def\cal{\mathcal}

\newcommand\fverb{\setbox\fverbbox=\hbox\bgroup\verb}
\newcommand\fverbdo{\egroup\medskip\noindent%
			\fbox{\unhbox\fverbbox}\ }
\newcommand\fverbit{\egroup\item[\fbox{\unhbox\fverbbox}]}
\newbox\fverbbox

\def\bea{\begin{eqnarray}}
\def\eea{\end{eqnarray}}
\def\be{\begin{equation}}
\def\ee{\end{equation}}

\def \del{\partial}

\newcommand\restr[2]{{
  \left.\kern-\nulldelimiterspace 
  #1 
  \vphantom{\big|} 
  \right|_{#2} 
  }}

\usepackage{accents}
\newlength{\dhatheight}

\usepackage{dsfont} 

\newcommand{\mrm}[1]{{\mathrm #1}}

\def\eps{\epsilon}
\def\phi{\varphi}

\newcommand{\reef}[1]{(\ref{#1})}

\def\eps{\epsilon}
\newcommand{\beq}{\begin{equation}} 
\newcommand{\eeq}{\end{equation}}
\def\del {\partial} 
\def\nn{\nonumber}

\def\calO {{\cal O}}

\def\calA {{\cal A}} 
\def\calB {{\cal B}} 
 
\def\calC {{\cal C}}

\def\half{{\textstyle\frac 12}}

\def\ge{\geqslant}

\newcommand{\diffop}[2]{\ifthenelse{\equal{#2}{1}}{\frac{\mrm{d}}{\mrm{d} #1}}{\frac{\mrm{d}^#2}{\mrm{d} #1^#2}}}

\newcommand{\NO}[1]{{:\!#1\!:}}

\usepackage[normalem]{ulem} 

\numberwithin{equation}{section}
\begin{document}

\vspace*{-.6in} \thispagestyle{empty}
\begin{flushright}
CERN PH-TH/2015-104\\
\end{flushright}
\vspace{1cm} {\Large
\begin{center}
{\bf The $\eps$-Expansion from Conformal Field Theory}\\
\end{center}}
\vspace{1cm}
\begin{center}
{\bf Slava Rychkov$^{a,b,c}$, Zhong Ming Tan$^{d}$, }\\[2cm] 
{
\small
$^{a}$ CERN, Theory Group, Geneva, Switzerland\\
$^{b}$ Laboratoire de Physique Th\'{e}orique de l'\'{E}cole Normale Sup\'{e}rieure (LPTENS), Paris, France
\\
$^{c}$ Facult\'e de Physique, Universit\'{e} Pierre et Marie Curie (UPMC), Paris, France 
\\
$^{d}$ D\'epartement de Physique, \'{E}cole Normale Sup\'{e}rieure, Paris, France
\normalsize
}
\\
\vspace{1cm}
\end{center}

\vspace{4mm}

\begin{abstract}
Conformal multiplets of $\phi$ and $\phi^3$ recombine at the Wilson-Fisher fixed point, as a consequence of the equations of motion.
Using this fact and other constraints from conformal symmetry, we reproduce the lowest nontrivial order results for the anomalous dimensions of operators, without any input from perturbation theory.
 \end{abstract}
\vspace{.2in}
\vspace{.3in}
\hspace{0.7cm} May 2015

\newpage

{
\setlength{\parskip}{0.05in}
\renewcommand{\baselinestretch}{0.7}\normalsize
\tableofcontents
\renewcommand{\baselinestretch}{1.0}\normalsize
}

\setlength{\parskip}{0.1in}

\section{Introduction}

The conformal bootstrap---using conformal symmetry to constrain the dynamics of various physically interesting critical theories---is an old idea 
\cite{Ferrara:1973yt,Polyakov:1974gs,Belavin:1984vu} 
which in the last few years has been shown to be much more powerful than previously thought \cite{Rattazzi:2008pe,Rychkov:2009ij,Caracciolo:2009bx,Poland:2010wg,Rattazzi:2010gj,Rattazzi:2010yc,Vichi:2011ux,Poland:2011ey}. A flagship result of this line of research is the world's most precise determination of the operator dimensions in the critical 3d Ising model \cite{Simmons-Duffin:2015qma}.\footnote{See \cite{Rychkov:2011et,ElShowk:2012ht,ElShowk:2012hu,El-Showk:2014dwa,Kos:2014bka} for prior work and \cite{Gliozzi:2013ysa,Gliozzi:2014jsa,Gliozzi:2015qsa} for closely related results by another technique.} 
Many other conformal theories, such as the $O(N)$ model \cite{Kos:2013tga,Kos:2015mba}, 
models with supersymmetry in various dimensions 
\cite{Beem:2013qxa,Bashkirov:2013vya,Alday:2013opa,Berkooz:2014yda,Alday:2014qfa,Chester:2014fya,Chester:2014mea,Beem:2014zpa,Bobev:2015vsa,Bobev:2015jxa}, 
as well as other less familiar or conjectural conformal theories
\cite{Gaiotto:2013nva,Nakayama:2014lva,Nakayama:2014yia,Nakayama:2014sba,Golden:2014oqa,Bae:2014hia,Chester:2014gqa} 
are being studied. 
In the hectic excitement of exploring a new research field, much of the recent development has been 
numerical.\footnote{Nevertheless, there have been notable analytic results for the large $N$ \cite{Heemskerk:2009pn,Heemskerk:2010ty,Fitzpatrick:2011dm}, large $\Delta$ \cite{Pappadopulo:2012jk,Goldberger:2014hca}, large $l$ \cite{Fitzpatrick:2012yx,Komargodski:2012ek,Fitzpatrick:2014vua,Vos:2014pqa,Kaviraj:2015cxa,Alday:2015eya,Kaviraj:2015xsa}, or SUSY \cite{Beem:2013sza,Beem:2014kka} cases. There is also a growing literature on the analytic expressions for the conformal correlation functions and conformal blocks, which is indispensable for the practical conformal bootstrap computations \cite{Dolan:2000ut,Dolan:2003hv,Fortin:2011nq,Costa:2011mg,Costa:2011dw,Dolan:2011dv,
Goldberger:2011yp,SimmonsDuffin:2012uy,Osborn:2012vt,Goldberger:2012xb,Hogervorst:2013sma,Fitzpatrick:2013sya,Hogervorst:2013kva,Dymarsky:2013wla,Fitzpatrick:2014oza,Behan:2014dxa,Khandker:2014mpa,Li:2014gpa,Costa:2014rya,Elkhidir:2014woa}.} 
It is a significant open problem to explain analytically the \emph{unreasonable effectiveness} of the numerical bootstrap.

An analytic conformal bootstrap attack on the 3d Ising model looks difficult, since one has to deal with a totally isolated conformal field theory (CFT). The $d=4-\eps$ case might be an easier target. Indeed, in the limit $\eps\to0$ the CFT should become close to the free theory. 
One can then hope to find a solution of the bootstrap equations as a power series in $\eps$.

The critical interacting scalar theory in $4-\eps$ dimensions is known as the Wilson-Fisher (WF) fixed point \cite{Wilson:1971dc}. 
It is normally studied using the standard diagrammatic perturbation theory and the renormalization group. 
This method also yields an expansion of observables, e.g.~the anomalous dimensions of local operators, as power series in $\eps$, called ``$\eps$-expansion'' series.
An interesting open question is whether conformal bootstrap can reproduce these series and perhaps even extend them to a higher order.\footnote{Numerically, conformal bootstrap equations in fractional dimensions were studied in \cite{El-Showk:2013nia}, matching the $\eps$-expansion series.}  

The purpose of this note is to report some progress on the last question. Using CFT methods, we will be able to reproduce the leading
$O(\eps)$ terms in the anomalous dimension of an infinite series of operators. For the lowest dimension scalar operator $\phi$ we will reproduce both $O(\eps)$ and $O(\eps^2)$ terms.  This may seem not much, given that the perturbative $\eps$-expansion series are known in some cases as far as $\eps^5$ \cite{Kleinert:1991rg} or even $\eps^6$ \cite{Kompaniets}. However, we believe that our discussion will lay the basis for any future extension to higher orders. Also, the modesty of the result is  compensated by the relative simplicity of the derivation.

The paper is organized as follows. In section \ref{sec:setup} we review the main structural properties of the WF fixed point. These properties are arrived at from the perturbative perspective, but we then formalize them in a series of ``axioms'' about CFT operators. The actual computation of operator dimensions, in section \ref{sec:computation}, proceeds from the axioms without recourse to the standard perturbation theory nor in fact to the Lagrangian.
In section \ref{sec:extension} we extend our computation is several directions, in particular generalizing to the 
$O(N)$ model. In section \ref{sec:open} we discuss open problems. Some technical details are relegated to appendix \ref{sec:confope}.

\section{Axioms}
\label{sec:setup}

To set the notation, consider the massless $\phi^4$ theory in $d=4-\eps$ dimensions (see e.g.~\cite{Kleinert:2001ax}):
\beq
S=\int d^dx \Bigl[\frac 12 (\del \phi)^2+\frac 1{4!} g\mu^\eps \phi^4\Bigr]\ .
\eeq
We will be interested in the IR fixed point of this theory, called the WF fixed point. It corresponds to the nontrivial zero of the $\beta$-function:
\beq
\beta(g)=-\eps g+ \frac{3}{16\pi^2} g^2 +O(g^3),
\eeq
which occurs for
\beq
g=g_*=\frac{16\pi^2}{3} \eps+O(\eps^2).
\eeq
Using the standard Feynman diagram perturbation theory, correlation functions of the $\phi^4$ theory can be expanded in the coupling constant. Renormalization group can then be used to reach the WF fixed point in the IR. All properties of the WF fixed point, such as the anomalous dimensions of local operators, are thus computable as series in $\eps$. 

Here, we will focus on the operators of the form $\phi^n$. As is well known, the anomalous dimension of $\phi$ arises first at two loops and is given by
\beq
\gamma_{\phi}=\frac{\eps^2}{108} +O(\eps^3)\,,
\label{eq:g1}
\eeq
while the higher powers of $\phi$ acquire anomalous dimensions already at one loop:\footnote{This result is usually given for $n=1,2,4$, but generalization to arbitrary $n$ is truly straightforward. }
\begin{gather}
\gamma_{\phi^n}=\frac 16 n(n-1)\eps +O(\eps^2),\quad n>1\,.
\label{eq:gn}
\end{gather}
As a matter of fact, the anomalous dimensions of $\phi$, $\phi^2$ and $\phi^4$ are known up to $\eps^5$ \cite{Kleinert:1991rg}, but we will not need those expressions here.

Our main result will be to present an alternative method of computing the anomalous dimensions of $\phi^n$. Instead of the diagrammatic perturbation theory, our method will be based on conformal field theory. By this method, we will be able to reproduce Eqs.~\reef{eq:g1},\reef{eq:gn} to the shown order. 

We will formalize our method by stating a series of three axioms. The axioms themselves can be motivated and justified from perturbation theory. However, once the axioms are agreed upon, the computations will be done by CFT techniques, without further recourse to the Lagrangian.

{\bf Axiom I} \emph{The WF fixed point is conformally invariant.} 

This can be proven to all orders in perturbation theory by considering the Ward identities of the renormalized stress tensor operator.\footnote{Conformal invariance follows from Eq.~(4.16) in \cite{Brown:1979pq} by specializing to $m^2=0$ and going to the IR where the $\beta$-function vanishes.} 

{\bf Axiom II} \emph{Correlation functions of operators at the WF fixed point approach free theory correlators in the limit $\eps\to0$.} 

This is pretty self-evident, since the fixed point coupling $g_*=O(\eps)$. A more concrete formulation of the same axiom is as follows:

{\bf Axiom II$\mathbf{'}$} \emph{For every local operator $\calO_{\rm free}$ in the free theory in $d=4$, there exists a local operator at the WF fixed point, $\calO_{{\rm WF}}$, which tends to $\calO_{\rm free}$ in the limit $\eps\to0$:} 
\beq
\lim _{\eps\to0}\calO_{{\rm WF}}= \calO_{{\rm free}}\,.
\eeq
The limit is understood in the sense of correlators, i.e.
\beq
\lim _{\eps\to0}\ \langle \calO_{{\rm WF}}(x_1) \calO'_{{\rm WF}}(x_2)\ldots \rangle_{d=4-\eps}= \langle \calO_{{\rm free}}(x_1) \calO'_{{\rm free}}(x_2)\ldots \rangle_{d=4}\,.
\eeq
{In particular, the scaling dimensions of WF operators must approach the free dimensions as $\eps\to 0$:\footnote{{Scaling dimensions of operators will be denoted by $[\calO]$ or $\Delta_\calO$.}}
\beq
\lim _{\eps\to0}\ [\calO_{{\rm WF}}] = [\calO_{{\rm free}}]\,.
\eeq
In agreement with what perturbation theory suggests, we will assume that scaling dimensions have a power series expansion in integer powers in $\eps$. In this paper we will consider these expansions as formal and will not be concerned with their convergence properties. It is in fact well known that the $\eps$-expansion series are not convergent but merely asymptotic \cite{Brezin:1976vw}. Moreover, the range of $\eps$ where the lowest-order result \reef{eq:gn} can be trusted presumably becomes smaller and smaller for larger $n$, due to the growth of the coefficient. This feature of the the $\eps$-expansion is also well known \cite{Kehrein:1992fn}, and we have nothing to add here.}
 
 Axioms I,II by themselves would not be sufficient to constrain the WF fixed point. This is because correlators of the free scalar theory in $d=4-\eps$ provide a trivial solution to both axioms. We need another axiom to distinguish the WF fixed point from the free theory.
This axiom can be inferred from the following reasoning. In the free theory all powers $\phi^n$ are independent local operators, with no derivative relations among them. From the CFT point of view, all these operators are what are called \emph{primaries}. This is not so at the WF fixed point, where we have one nontrivial relation ($\Box\equiv \del^2$):
\beq
\Box \phi=\frac 1{3!} g_*\mu^\eps \phi^3\,.
\label{eq:eom}
\eeq
This is nothing but the classical equation of motion, and it survives in the quantum theory for renormalized operators. Consequently, at the WF fixed point, the $\phi^3$ is not an independent operator, but is a derivative of $\phi$. In CFT, such operators are called \emph{descendants},
and their properties are completely fixed in terms of those of a primary. In particular, the total dimension (classical plus anomalous) of $\phi^3$ is predicted to be two plus that of $\phi$:
\beq
\Delta_{\phi^3}= \Delta_{\phi}+2\qquad\text{(WF)}\,.
\label{eq:31WF}
\eeq 
This relation has to hold to all orders in $\eps$, and it can be verified to $O(\eps)$ using Eqs.~\reef{eq:g1},\reef{eq:gn}.

Once $\phi^3$ is recognized as a descendant of $\phi$, its descendants also become descendants of $\phi$. A primary operator together with all its descendants constitute a \emph{conformal multiplet}. We thus observe the phenomenon of multiplet recombination---two distinct multiplets in the free theory---those of $\phi$ and $\phi^3$---join and become a single conformal multiplet of $\phi$ at the WF fixed point:
\beq
\{\phi\}_{\rm WF}\approx \{\phi\}_{\rm free} + \{\phi^3\}_{\rm free}\,.
\label{eq:rec}
\eeq
Here, $\{\ldots\}$ denotes a conformal multiplet, and $\approx$ means that the number of states at each level is conserved in the process of recombination, up to small anomalous dimensions. 

Another way to see the need for multiplet recombination is as follows. The free scalar field satisfies the equation $\Box\phi_{\rm free}=0$,  which acts as a shortening condition for the multiplet $\{\phi\}_{\rm free}$, setting many descendants to zero. There is no such shortening condition for the multiplet $\{\phi\}_{\rm WF}$, which thus contains more states that $\{\phi\}_{\rm free}$. These new states must come from some other multiplet. One can see that $\{\phi^3\}_{\rm free}$ contains the right number of states with the right dimensions to supply the difference.\footnote{This can be also seen from the character formulas for the corresponding representations of the conformal group \cite{Barabanschikov:2005ri}.} Moreover, the free theory contains no other candidate multiplet able to fulfil this role. The recombination pattern in Eq.~\reef{eq:rec} is thus mandatory already from the point of view of counting states. 

The just described phenomenon of multiplet recombination can be viewed as a sort of CFT analogue of the familiar Higgs mechanism from particle physics. Just as a massless vector boson can acquire mass only by ``eating'' a scalar Goldstone particle, which provides a necessary extra state, so a short conformal multiplet can become long only by eating another multiplet.

Notice that $\phi$ and $\phi^3$ are the only free scalar primaries whose multiplets recombine.\footnote{\label{note:spinrec}As will be discussed in section \ref{sec:open}, recombination also happens for the multiplets of conserved currents of spins $l>2$. In holographic theories, the Higgs mechanism analogy becomes literal for this recombination, as it is described via the usual Higgs mechanism in the bulk, see e.g.~\cite{Bianchi:2005ze}. We are grateful to Leonardo Rastelli for emphasizing the last point to us.} The multiplets of $\phi^2$ and of $\phi^n$ ($n\ge4$) are already long in the free theory, and they remain long in the interacting theory.

From now on, we will adhere to a stricter notation for operators. The notation $\phi^n$ will be used to denote exclusively the free theory operators, while the WF operators $(\phi^n)_{\rm WF}$ will be denoted by $V_n$. Operators $V_n$ and $\phi^n$ are related in the sense of Axiom II$'$:
\beq
\lim_{\eps\to 0} V_n =\phi^n\,.
\label{eq:vn}
\eeq
Our last axiom summarizes the above discussion about the nature of the operators $V_n$. It will allow us to distinguish the WF fixed point from the free theory in $4-\eps$ dimensions.

{\bf Axiom III} \emph{Operators $V_n$, $n\ne 3$, are primaries. Operator $V_3$ is not a primary but is proportional to $\Box V_1$:}
\beq
\Box V_1 =  \alpha V_3 \,.
\label{eq:axIII}
\eeq
The proportionality coefficient $\alpha=\alpha(\eps)$ should be considered as an unknown at this stage. It would be contrary to our philosophy to fix it using the equation of motion \reef{eq:eom}. Our logic is that perturbation theory should be used to infer only robust, structural properties, like the fact that multiplets recombine. The actual computations should be done using CFT. Below we will be able to determine $\alpha$ by imposing Eq.~\reef{eq:vn}. 

\section{Computation}
\label{sec:computation}

As already said, we will be interested in the dimensions of the operators $V_n$, which can be written as
\beq
\Delta_n = n \delta +\gamma_n\,.
\eeq
Here $\delta=1-\eps/2$ is the free scalar dimension in $4-\eps$ dimensions, and $\gamma_n$ is the anomalous dimension, which can be expanded in $\eps$:
\beq
\gamma_n = y_{n,1} \eps + y_{n,2} \eps^2+\ldots
\eeq
{As mentioned above, we will assume that only integer powers of $\eps$ occur in this expansion.}

Perturbation theory results in Eqs.\reef{eq:g1},\reef{eq:gn} are equivalently stated as:
\begin{gather}
y_{n,1}=\frac 16 n(n-1)\qquad(n=1,2,\ldots),\qquad\qquad y_{1,2}=1/108\,.
\label{eq:toprove}
\end{gather}
In this section, we will show that these same results can be derived from the axioms using CFT reasoning.

The key idea is to consider correlators involving operators $V_n$ and $V_{n+1}$:
\beq
\langle V_n (x_1) V_{n+1}(x_2) \ldots \rangle\,, 
\label{eq:corrWF}
\eeq
where $\ldots$ stands for some other operator insertions. In the limit $\eps\to0$ these correlators should approach 
\beq
\langle \phi^n (x_1) \phi^{n+1}(x_2) \ldots \rangle\,.
\label{eq:corrfree}
\eeq
These correlators will turn out to be useful because of their sensitivity to the multiplet recombination phenomenon.
In the free theory the operator product expansion (OPE) $\phi^n\times \phi^{n+1}$ contains both operators $\phi$ and $\phi^3$, with coefficients independently computable by counting Wick contractions. On the other hand, in the interacting theory the OPE $V_n\times V_{n+1}$ contains only $V_1$ as a primary, while $V_3$ appears in the guise of $\Box V_1$, with a relative coefficient fixed by conformal symmetry. A nontrivial consistency condition will then arise from the requirement that the $\eps\to0$ limit of the second OPE should agree with the first one.

Let's supply the details. It will be convenient to change the normalization of the free 4d scalar: 
\beq
\phi_{\rm new}=2\pi \phi_{\rm old}\,.
\label{eq:ch}
\eeq
 In the new normalization the two point function of $\phi$ is unit-normalized:
\beq
\langle \phi(x)\phi(0)\rangle = 1/|x|^2\,.
\eeq
Normalization of $V_1$ can then be chosen so that
\beq
\langle V_1(x) V_1(0)\rangle = 1/|x|^{2\Delta_1}\,.
\label{eq:v1v1}
\eeq
Axiom II$'$ is satisfied. 

Our first task is to find $\alpha$ in Axiom III. By repeated differentiation of \reef{eq:v1v1}, we get:
\begin{gather}
\langle \Box V_1(x) \Box V_1(0)\rangle = h/|x|^{2\Delta_1+4}\,,\\[5pt]
h=16\Delta_1(\Delta_1+1)(\Delta_1-\delta)(\Delta_1+1-\delta)\approx 32\gamma_1\qquad (\eps\ll1)\,.
\end{gather}
On the other hand we must have
\beq
\langle V_3(x) V_3(0)\rangle \to \langle \phi^3(x) \phi^3(0)\rangle =6/|x|^6\,.
\eeq
Comparing the prefactors of the last two equations, we determine $\alpha$ up to a sign:
\beq
\alpha=4 \sigma (\gamma_1/3)^{1/2}\,,\qquad \sigma=\pm 1\,.
\label{eq:alpha}
\eeq
Later on, we will be able to fix the sign ambiguity and show that $\sigma=1$.

Now, let's consider the correlators of the form \reef{eq:corrWF} and \reef{eq:corrfree}. It will be sufficient to analyze them in the region where the points $x_1$ and $x_2$ are much closer to each other than to the other operator insertions, so that we can use the OPE.

We will not need the whole OPE, but only its leading part sensitive to the multiplet recombination. In the 4d free theory these are the $\phi$ and $\phi^3$ terms:\footnote{All local operators in the free theory are assumed normal ordered.}
\beq
\phi^n(x) \times \phi^{n+1}(0) \supset f {|x|^{-2n}}\bigl\{\phi(0) + \varrho {|x|^{2}}\phi^3(0)\bigr\} \,.
\label{eq:OPEfree}
\eeq
The OPE coefficients are found by counting the number of independent Wick contractions:\footnote{{This involves some combinatorics. One may wonder if the calculation of these and similar constants below can be simplified using the conformal symmetry (or the higher spin symmetry) of the free scalar theory, but we have not explored this here.}} 
\beq
f =(n+1)!,\qquad \varrho= n/2\,.
\label{eq:N=1}
\eeq
The needed terms in the OPE at the WF fixed point are:
\beq
V_n(x) \times V_{n+1}(0) \supset \tilde f |x|^{\Delta_1-\Delta_n-\Delta_{n+1}} (1 + q_1 x^\mu \del_\mu
+ q_2 x^\mu x^\nu \del_\mu \del_\nu + q_3 x^2 \Box+\ldots)V_1(0)\,.
\label{eq:OPEWF}
\eeq

We will now consider the following two correlators in the free theory:
\beq
\langle \phi^n (x) \phi^{n+1}(0) \phi(z)\rangle\text{  and  } \langle \phi^n (x) \phi^{n+1}(0) \phi^3(z)\rangle\,,
\label{eq:twocorr}
\eeq
and the WF correlators 
\beq
\langle V_n (x) V_{n+1}(0) V_1(z)\rangle \text{  and  }\langle V_n (x) V_{n+1}(0) V_3(z)\rangle\,,
\eeq
which have to tend to their counterparts in \reef{eq:twocorr} for $\eps\to0$. In the configuration $|x|\ll|z|$ the leading behavior of the free correlators is found using the above OPE and is given by:
\begin{gather}
\langle \phi^n (x) \phi^{n+1}(0) \phi(z)\rangle\approx {f }{|x|^{-2n}}\langle \phi(0) \phi(z)\rangle\,,
\label{eq:free1}\\[7pt]
\langle \phi^n (x) \phi^{n+1}(0) \phi^3(z)\rangle\approx {f \varrho}{|x|^{-2n+2}}\langle \phi^3(0) \phi^3(z)\rangle\,.
\label{eq:free2}
\end{gather}
Using the WF OPE, we also have the behavior of the WF correlators in the same configuration. For the first correlator we have
\beq
\langle V_n (x) V_{n+1}(0) V_1(z) \rangle \approx \tilde f |x|^{\Delta_1-\Delta_n-\Delta_{n+1}} \langle V_1(0) V_1(z)\rangle\,.
\eeq
This will match the free correlator if $\tilde f \approx f $ for $\eps\to0$, and consequently $\tilde f$ will be non-vanishing.
The second correlator is more interesting. We have:
\begin{align}
\langle V_n (x) V_{n+1}(0) V_3(z) \rangle \approx & \tilde f |x|^{\Delta_1-\Delta_n-\Delta_{n+1}} \nn\\
&\times(1 + q_1 x^\mu \del_\mu
+ q_2 x^\mu x^\nu \del_\mu \del_\nu + q_3 x^2 \Box+\ldots) \langle V_1(0) V_3(z)\rangle\,,
\label{eq:WF2}
\end{align}
where the derivatives fall on the argument of $V_1(0)$ inside the two point function. The coefficients $q_i$ are fixed by conformal symmetry in terms of the dimensions of the involved operators. Their expressions are computed in appendix \ref{sec:confope} and will be discussed below. 

We have to match \reef{eq:WF2} with the free correlator \reef{eq:free2}. First of all, notice that \reef{eq:free2} predicts the subleading behavior $O(x^2)$ with respect to \reef{eq:free1}. However, in \reef{eq:WF2} we see terms $O(1)$ as well as $\sim x_\mu$ and $\sim x_\mu x_\nu$. For the matching to occur, all these offending terms have to vanish in the limit $\eps\to0$. In fact, using Eqs.~\reef{eq:axIII} and \eqref{eq:alpha} we obtain:
\beq
\langle V_1(0) V_3(z)\rangle=\frac {4\Delta_1(\Delta_1-\delta) \alpha^{-1}}{|z|^{2\Delta_1+2}} = \frac{\sigma\Delta_1\sqrt{3\gamma_1}}{|z|^{2\Delta_1+2}}\,,
\eeq
which goes to zero for $\eps\to0$ as $\sqrt{\gamma_1}$. Therefore, the above-mentioned offending terms will go away as long as $q_1,q_2$ remain finite in the $\eps\to0$ limit. 
We will see below that this condition is indeed satisfied. 

On the other hand, to match the terms $O(x^2)$, we must have that $q_3$ \emph{blows up} in the $\eps\to0$ limit at a rate which can be precisely determined. Namely, we must have:\footnote{The coefficient of $x^2$ also gets a contribution from the $q_2$ term in the OPE. This extra contribution can be ignored in Eq.\reef{eq:qa} because, as mentioned, $q_2$ will remain finite for $\eps\to0$, while $q_3$ will blow up.}
\beq
q_3\alpha \to \varrho=n/2\qquad (\eps\to0)\,.
\label{eq:qa}
\eeq
The rest of the argument will follow from Eqs.~\reef{eq:alpha}, \reef{eq:qa} and the explicit expressions for the coefficient $q_3$.

As already mentioned, $q_i$ are fixed by conformal symmetry in terms of the dimensions of the involved operators. 
The full expressions will be given in appendix \ref{sec:confope}. We have to consider separately two cases: 
\begin{alignat}{3}
&\text{(i)}\ \ n&=1&\text{ and }n\ge 4\,,\nn\\
&\text{(ii)}\ n&=2&\text{ and }n=3\,.
\label{eq:twocases}
\end{alignat}
In case (i), both operators in the LHS of the OPE, $V_n$ and $V_{n+1}$, are primaries. In case (ii), one of them is the descendant $V_3$.
Since descendants transform under the conformal group differently from primaries, their OPE has to be derived separately.
 
Let us begin by considering case (i). The coefficients $q_i$ are then given by Eqs.~\reef{eq:q3prim}, where we have to put $a=\Delta_n$, $b=\Delta_{n+1}$, $c=\Delta_1$. It's easy to see that $q_1$, $q_2$ remain nonsingular as $\eps\to0$, in agreement with the above discussion. On the other hand, $q_3$ has a chance to become singular because of the $c-\delta$ factor in the denominator. The asymptotic expression for $q_3$ in this limit is given by: 
\beq
q_3\approx \frac{\gamma_{n+1}-\gamma_n+\gamma_1}{16\gamma_1}\qquad  (\eps\ll1).
\label{eq:q}
\eeq
The numerator in this formula is at least $O(\eps)$. To have a singularity as $\eps\to0$, the $\gamma_1$ in the denominator must be at least $O(\eps^2)$, i.e.
\beq
y_{1,1}=0\,.
\eeq
Let us assume for the moment that $\gamma_1$ is precisely $O(\eps^2)$, i.e.
\beq
\gamma_1\approx y_{1,2} \eps^2\,,\qquad y_{1,2}\ne0\,.
\eeq
We will be able to justify this assumption later in this section.  
Substituting this asymptotics for $\gamma_1$ into Eqs.~\reef{eq:alpha},\reef{eq:q}, we obtain:
\begin{gather}
\alpha\approx4 \sigma (y_{1,2}/3)^{1/2}\eps \,,\label{eq:a1}\\
q_3\approx (y_{n+1,1}-y_{n,1})/(16 y_{1,2}\eps)\,.
\label{eq:q3}
\end{gather}
Using these expressions in \reef{eq:qa} gives:
\beq
y_{n+1,1}-y_{n,1}=2\sigma (3 y_{1,2})^{1/2} n\,.
\label{eq:n+1,n}
\eeq

So far we have considered only case (i). We now turn to case (ii), which involves operator $V_3$ on the LHS of the OPE. Operator $V_3$ is a descendant of $V_1$ and its OPE can be worked out starting from the OPE of $V_1$. This computation is done in appendix \ref{sec:confope}. The result for $q_3$ can be expressed in terms of $\Delta_1$ and the dimension of the third operator ($V_2$ or $V_4$). Substituting $\gamma_1\approx y_{1,2} \eps^2$, the result simplifies and we find (see Eqs.\reef{eq:n2},\reef{eq:n3} where $q_3$ was denoted $\tilde q_3$):
\beq
q_3\approx \begin{cases}(1-y_{2,1})/(16y_{1,2}\eps)&(n=2)\,,\,\\
(y_{4,1}-1)/(16y_{1,2}\eps)&(n=3)\,.
\end{cases}
\label{eq:q3ii}
\eeq

Now, notice that
\beq
y_{3,1}=1\,,
\eeq 
which follows from $y_{1,1}=0$ and from the fact that $\Delta_3-\Delta_1=2$ (see \reef{eq:31WF}). Using this fact, we can restate Eqs.\reef{eq:q3ii} as:
\beq
q_3\approx (y_{n+1,1}-y_{n,1})/(16 y_{1,2}\eps),\qquad (n=2,3)\,.
\eeq
Amazingly, this is the same equation as the one we found for $n=1$ and $n\ge 4$ in Eq.~\reef{eq:q3}. Thus we conclude that Eq.~\reef{eq:q3} is in fact true for all $n\ge 1$.

This allows us to complete the computation. Eq.~\reef{eq:n+1,n} was a consequence of \reef{eq:q3} and we can now say that it's true for all $n\ge 1$. Solving the recursion \reef{eq:n+1,n} with the initial condition $y_{1,1}=0$, we find
\beq
y_{n,1}=K n(n-1)\,,
\label{eq:yn}
\eeq
where
\beq
K= \sigma(3 y_{1,2})^{1/2}\,.
\label{eq:K}
\eeq
The coefficient $K$ is determined by imposing that $y_{3,1}=1$, which gives $K=1/6$. 
Using the relation \reef{eq:K} we find $\sigma=+1$ and $y_{1,2}=1/108$. Eq.~\reef{eq:toprove}
has now been fully derived as promised.

As a side remark, it's reassuring that the value of the $\alpha$ coefficient implied by \reef{eq:a1} can now be seen to agree with the proportionality coefficient $g_*/3!$ in the perturbative equation of motion \reef{eq:eom}, once the change in the field normalization \reef{eq:ch} is taken into account.

It remains to justify the assumption we made above that $y_{1,2}\ne0$. Consider Eq.~\reef{eq:q} for $n=1$:
\beq
q_3\approx \gamma_2/(16 \gamma_1)\qquad(n=1)\,.
\eeq
Since by Eq.~\reef{eq:qa} this must go as $1/\alpha\sim 1/\sqrt{\gamma_1}$ for $\eps\to 0$, we conclude that 
\beq
\gamma_1\sim(\gamma_2)^2\,.
\label{eq:g11}
\eeq
Consider further the same coefficient for $n=2$, whose asymptotics was given in appendix \ref{sec:confope}, Eq.~\reef{eq:n2gen}:
\beq
q_3\approx (\eps-\gamma_2)/(16 \gamma_1)\qquad(n=2)\,.
\eeq
This also must go as $1/\alpha\sim 1/\sqrt{\gamma_1}$, and hence 
\beq
\gamma_1\sim(\eps-\gamma_2)^2\,.
\label{eq:g12}
\eeq
This equation can hold together with \reef{eq:g11} only if $\gamma_1\sim \eps^2$, i.e.~if $y_{1,2}\ne0$. The derivation is now complete.

Looking at the above computation, one can't help but wonder \emph{why} Eq.~\reef{eq:q3}, initially derived for the all primary case (i), eventually turned out to be also true when one of the involved operators is $V_3$, a descendant. An intuitive reason for why this happened is as follows. $V_3$ is not just any descendant, but a descendant which is on the verge of becoming null\footnote{An operator is called null if it has a zero two point function with itself.}(it becomes null when $\gamma_1\to0$). It's a fact familiar from 2d CFT that the leading null descendant in a conformal multiplet behaves like a primary.

One concrete relevant example of this phenomenon is as follows. Consider the correlator of three scalar primary operators $\calA$, $\calB$, $\calC$ of dimensions $a,b,c$, which is fixed by conformal symmetry to have the form \cite{Polyakov:1970xd}: 
\beq
\langle \calA(x) \calB(y) \calC(z)\rangle = f |x-y|^{c-a-b}|x-z|^{b-a-c}|y-z|^{a-b-c}\,.
\label{eq:3pt}
\eeq
Now apply $\Box_x$ to this equation, to compute the correlator
\beq
\langle \Box\calA(x) \calB(y) \calC(z)\rangle\,. 
\label{eq:3ptBox}
\eeq
Even though this is also a correlator of three scalars, for general $a$,$b$,$c$ it will not be of the form \reef{eq:3pt}.
This is not surprising since $\Box\calA$ is not a primary. However, let's now consider the limit $a\to \delta$, which is precisely the limit when $\Box\calA$ becomes null. It's easy to check that in this limit \reef{eq:3ptBox} takes the form of \reef{eq:3pt} with $a=\delta+2$
and $\tilde f = (\delta+b-c)(\delta+c-b) f$. 

So, to summarize, the intuitive reason why Eq.~\reef{eq:q3} turns out to be valid also for $n=2,3$ is that $V_3$, though not a primary, is almost null and is behaving almost like a primary. It may be possible to derive Eq.~\reef{eq:q3} for $n=2,3$ by quantifying the ``almost" in the previous sentence. However, in our presentation we found it easier to do an explicit computation.

\section{Extensions}
\label{sec:extension}

We will now discuss a few simple extensions of the calculation from the previous section. The main idea is the same, so we will be brief.

\subsection{Other primary scalars}

Up to now, we focussed on the scalar operators of the form $\phi^n$. The crucial idea was to consider the OPE $\phi^{n}\times \phi^{n+1}$, which contains the operator $\phi$. Here we will indicate how the same argument can be applied to other OPEs containing $\phi$.

So, let $\calA$ be a scalar primary\footnote{Recall that the primary operators, by definition, satisfy the condition $[K_\mu,\calO(0)]=0$, where $K_\mu$ is the special conformal transformation generator.} operator in the free massless 4d scalar theory, and let $\calB$ be the normal-ordered product of $\phi$ and $\calA$:
\beq
\calB=\NO{\phi \calA}\,.
\eeq
Then $\calB$ is also a primary, and the OPE of $\calA$ and $\calB$ contains $\phi$.\footnote{It can be shown generally that if the OPE of two scalar primaries contains $\phi$ then one of them is the normal-ordered product of $\phi$ with the other. We thank Balt van Rees for a discussion of this point.}
We denote by $\varrho$ the relative coefficient of the operator $\phi^3$ in the same OPE: 
 \beq
 \calA(x)\times \calB(0)\supset f|x|^{-[\calA]-[\calB]+1} \bigl\{\phi(0)+ \varrho\, x^2 \phi^3(0)\bigr\}\qquad(\text{free, } d=4)\,.\label{eq:ABfree}
\eeq
This is the counterpart of \reef{eq:OPEfree}. We assume that $f\ne 0$, while $\varrho$ may or may not be zero. 

We will denote by $\calA^\eps$, $\calB^\eps$, the operators at the WF fixed point which tend to $\calA$, $\calB$ in the sense of Axiom II$'$. 
The counterpart of \reef{eq:OPEWF} will be:
\beq
\calA^\eps(x)\times \calB^\eps(0)\supset \tilde f  |x|^{-[\calA^\eps] - [\calB^\eps]+\Delta_1} \bigl\{V_1(0)+ (q_3 \alpha) x^2 V_3(0)\bigr\}
\label{eq:ABWF}
\qquad(\text{WF, } d=4-\eps)\,.
\eeq
We kept only the terms in this OPE that will play a role in the subsequent discussion.

Repeating the argument of section \ref{sec:computation}, we can show that the coefficients of the subleading terms in \reef{eq:ABfree} and \reef{eq:ABWF} must match:
\beq
q_3\alpha \approx \varrho \qquad(\eps\to0)
\label{eq:tc1}
\eeq
On the other hand, we can determine $q_3$ using conformal symmetry.
We will assume that operators $\calA^\eps$, $\calB^\eps$ are primaries of the interacting CFT.\footnote{The only case when this does not happen is if one of them is $V_3$. But then the other operator is necessarily $V_2$ or $V_4$, and these cases were already treated in section \ref{sec:computation}.} The $q_3$ is then given by Eq.~\reef{eq:q3prim}. Its asymptotic behavior of $q_3$ in the $\eps\to 0$ limit can then be written as:
\beq
q_3\approx {([\calB^\eps]-[\calA^\eps]-\Delta_1)}/({16\gamma_1}) \,.
\label{eq:tc2}
\eeq
Here, we kept the dependence on $\eps$ only in the factors which have a chance to vanish for $\eps\to0$.
Comparing this equation with \reef{eq:tc1}, we get a condition which relates the considered dimensions at $O(\eps)$:
\begin{align}
[\calB^\eps]-[\calA^\eps]-\delta|_{d-4-\eps}&\approx 4\sqrt{3\gamma_1}\varrho=(2\varrho/3)\,\eps +O(\eps^2)\,.
\end{align}
This equation generalizes the recursion relation for the dimensions of the $\phi^n$ operators from
section \ref{sec:computation}. 

The above discussion is meant to demonstrate that there is further potential for applying the recombination to constrain operator dimensions. We could continue by considering explicit examples of scalar primary operators containing derivatives. Unfortunately, even the simplest such operators are a bit awkward to work with (they have 4 $\phi$'s and 4 derivatives). So we postpone full exploration of this topic to future work.

\subsection{Generalization to the $O(N)$ model}

We will now generalize to the $O(N)$ model. The fixed point for the $O(N)$ model is defined starting from the Lagrangian with $N$ massless scalar fields $\phi^a$ in $4-\eps$ dimensions and turning on the interaction $(\phi^a\phi^a)^2\equiv (\vec \phi^{\,2})^2$, which preserves the global $O(N)$ symmetry. The operators will form multiplets of the global symmery group. We will not attempt an exhaustive analysis, but will consider just two series of operators. 

Our first series will consist of, on the free theory side, operators
\begin{alignat}{3}
\Phi_{2p+1}^a\equiv \phi^a(\vec\phi^{\,2})^p&\qquad&\text{and}\qquad &\Phi_{2p}\equiv(\vec\phi^{\,2})^p\,.
\intertext{
The fixed point operators tending to them in the sense of Axiom II$'$ will be denoted as
}
W_{2p+1}^a &\qquad&\text{and}\qquad &W_{2p} \,,
\label{eq:Ws}
\end{alignat}
with their respective anomalous dimensions $\gamma_{2p+1}$ and $\gamma_{2p}$. 

The $W$ operators are primaries, except for $W_3^a$ which is a descendant of $W_1^a$:
\beq
\Box W_1^a = \alpha W_3^a\,.
\eeq
This relation is justified similarly to Axiom III of the WF fixed point for $N=1$. The discussion which led to Eq.~\reef{eq:alpha} readily generalizes, and gives
\beq
\alpha = 4\sigma [\gamma_1/(2+N)]^{1/2}\,\qquad \sigma=\pm 1\,.
\eeq

Eq.~\reef{eq:OPEfree} splits into two equations, depending whether $n$ is even or odd:
\begin{align}
\Phi_{2p}(x)\times \Phi_{2p+1}^a(0) &\supset f_{2p}|x|^{-4p}\bigl\{\Phi_1^a(0)+\rho_{2p}\, x^2\, \Phi_3^a(0)\bigr\}\,,\\
\Phi^a_{2p+1}(x)\times \Phi_{2p+2}(0) &\supset f_{2p+1}|x|^{-4p-2}\bigl\{\Phi_1^a(0)+\rho_{2p+1}\, x^2\, \Phi_3^a(0)\bigr\}\,.
\end{align}
We'll normalize $\phi$ so that the two point function is $\langle \phi^a(x) \phi^b(0)\rangle=\delta^{ab}/|x|^{2}$.
In this normalization, the relative coefficients in the above OPEs are given by:
\beq
\varrho_{2p} = \frac{3p}{2+N}\,,\qquad
\varrho_{2p+1} = \frac{3(2p+1)+N-1}{2(2+N)}\,.
\eeq
The overall coefficients are not needed below, but we give them for completeness:
\beq
 f_n = \prod_{k=2}^{n+1} w_k,\qquad w_{2p}=2p,\ w_{2p+1}=2p+N\,.
\eeq
For $N=1$ these formulas reduce to Eq.~\reef{eq:N=1}.

Now repeating the steps which led to Eq.~\reef{eq:n+1,n}, we obtain:
\beq
y_{n+1,1}-y_{n,1}=4\sigma[(N+2) y_{1,2}]^{1/2} \varrho_n\,.
\label{eq:O(N)rec0}
\eeq
Just as in section \ref{sec:computation}, this equation is valid also for $n=2,3$ even though $W_3^a$ is not a primary. 
Summing the equations for $n=1$ and $n=2$, we get:
\beq
y_{3,1}-y_{1,1}=4\sigma[(N+2) y_{1,2}]^{1/2}(\varrho_1+\varrho_2)\,.
\eeq
On the other hand, just as in section \ref{sec:computation}, we know that
\beq
y_{1,1}=0\,,\qquad y_{3,1}=1\,.
\label{eq:bdry}
\eeq
It follows that $\sigma=1$ and
\beq
y_{1,2}=\frac{N+2}{4(N+8)^2}\,.
\label{eq:g1O(N)}
\eeq
Substituting this into \reef{eq:O(N)rec0}, we obtain
\beq
y_{n+1,1}-y_{n,1}=\frac{2(N+2)}{N+8} \varrho_n\,.
\label{eq:O(N)rec}
\eeq
From here and \reef{eq:bdry}, all the unknown $y_{n,1}$'s can be found recursively.
E.g., we get
\beq
y_{2,1}=\frac{N+2}{N+8}\,,\qquad y_{4,1}=2\,.
\eeq
These results as well as Eq.~\reef{eq:g1O(N)} agree with diagrammatic methods (see e.g.~\cite{Kleinert:2001ax}).
The general solution of \reef{eq:O(N)rec} is
\begin{align}
y_{2p+1,1} = \frac{p(N+2+6p)}{N+8},\qquad y_{2p,1} = \frac{p[N+2+6(p-1)]}{N+8}\,,
\end{align}
in agreement with the diagrammatic result in Eq.~(60) of \cite{Kehrein:1992fn}.

The second series that we consider consists, on the free theory side, of the operators
\beq
\Phi_{2p}^{ab}\equiv [\phi^a \phi^b-N^{-1}\delta^{ab} \vec\phi^{\,2}](\vec\phi^{\,2})^{p-1}\,,
\eeq
transforming as symmetric traceless 2-tensors. The relevant OPE sensitive to the multiplet recombination is
\beq
\Phi_{2p-1}^{c} \times \Phi_{2p}^{ab} \supset |x|^{-4p+2}h_{p} \Gamma^{c|ab|e}\{\Phi_1^e+ \kappa_{p} x^2 \Phi_3^e\} \,,
\label{eq:OPEST}
\eeq
where $\Gamma$ is the Clebsch-Gordan coefficient for the corresponding $O(N)$ representations:
\beq
\Gamma^{c|ab|e}=\delta^{ac}\delta^{be}+ \delta^{bc}\delta^{ae}-(2/N)\delta^{ab}\delta^{ce}\ .
\eeq
The OPE coefficients are given by
\beq
h_p=\frac{1}{2p(2+N)}\prod_{k=1}^p 2k(2k+N)\,,\qquad \kappa_p=\frac{3p-2}{2+N}\,.
\eeq
Notice that to determine $\kappa_p$, the contribution of $\Phi_3^e$ has to be carefully disentangled from the contribution 
of the symmetric traceless 3-tensor operator
\beq
\phi^a\phi^b\phi^c - \frac{1}{2+N}\vec\phi^{\,2}(\delta^{ab}\phi^c+\delta^{ac}\phi^b+\delta^{bc}\phi^a)\,,
\eeq
which has the same scaling dimension.\footnote{Further details on this point and on other aspects of our paper can be found in \cite{report}.}
Applying the usual argument to Eq.~\reef{eq:OPEST} gives a relation between $O(\eps)$ anomalous dimensions of the involved operators at the WF fixed point:
\beq
y^{\rm ST}_{2p,1} - y_{2p-1,1}=\frac{2(N+2)}{N+8} \kappa_p\,.
\eeq
In particular, for the anomalous dimension of $\Phi_2^{ab}$ we get
\beq
y_{2,1}^{\rm ST}=\frac{2}{N+8}\,,
\eeq
which is a well-known result \cite{Wilson:1972cf}, while for general $p$ we get
\begin{align}
y_{2p,1}^{\rm ST} = \frac{N(p-1)+2p(3p-2)}{N+8}\,.
\end{align}
\section{History and prior work}
\label{sec:history}
This project had a somewhat long gestation period, and several other people contributed to it at the initial stages. We would like to acknowledge their contributions here.

The problem of determining the $\eps$-expansion series by CFT techniques is rather old. It was discussed already by Polyakov in 1974 \cite{Polyakov:1974gs}; a comparison to his results will be given below. 

This problem gained renewed attention at the ``Back to the Bootstrap II'' workshop (Perimeter Institute, June 2012), where the results of \cite{Liendo:2012hy} were reported. Among other things, that paper analyzed boundary conformal bootstrap equations in $4-\eps$ dimensions, for the conformal two point function $\langle \phi \phi\rangle$ in half-space with the Dirichlet or Neumann boundary conditions. In both cases, a one-parameter family of solutions of crossing symmetry constraints to $O(\eps)$ was found. All the solutions had $\gamma_1=O(\eps^2)$, and to pick one uniquely within the family, $\gamma_2$ had to be provided by some other means, e.g.~from perturbation theory.

At the same workshop, one of us (S.R.) learned about the multiplet recombination phenomenon from Leonardo Rastelli and Balt van Rees. 
Shortly afterwards, Sheer El-Showk, Miguel Paulos, S.R.~and David Simmons-Duffin found that multiplet recombination provides a constraint on the anomalous dimensions of $V_1$ and $V_2$. They looked at the conformal block of $V_1$ appearing in the decomposition
of the four point function $\langle V_1(x_1) V_2(x_2) V_1(x_3) V_2(x_4)\rangle$, using the series representation from \cite{Dolan:2000ut}, Eq.~(2.32). Demanding that this block reduce for $\eps\to0$ to the sum of the $\phi$ and $\phi^3$ blocks in the free theory, they arrived at a relation 
identical to the $n=1$ case of \reef{eq:n+1,n}. Then, borrowing $y_{1,1}=0$, $y_{2,1}=1/3$ from perturbation theory, they could determine $y_{1,2}$. This computation was later written up as a problem for the 2013 Mathematica School on Theoretical Physics \cite{MathProb,MathNb}. 

Compared to this previous work, our paper added two main ideas. First, we showed that the multiplet recombination can be exploited already at the level of three point functions, which is rather easier than to analyze the four point function and its conformal blocks. 
Second, we realized that one should study all $\phi^n\times \phi^{n+1}$ OPEs together. In particular, this links $\phi$ to $\phi^3$ in two steps. This eventually allowed us to determine all the unknown quantities without any input from perturbation theory. 

The pioneering computation of Polyakov \cite{Polyakov:1974gs}, section 5, deserves a separate comment. He analyzed the $O(N)$ model four point function $\langle \phi^a \phi^b \phi^c \phi^e \rangle$ using CFT methods. The computation was done using not conformal blocks, but the ``unitary blocks" which he introduced and which satisfy crossing symmetry but violate OPE by logarithmic terms.\footnote{These unitary blocks are closely related, if not identical, to the Witten diagrams of AdS/CFT \cite{Polyakov}.} 
His consistency condition was that these OPE-offending terms must cancel. He analyzed this condition to $O(\eps)$. Assuming that the anomalous dimension of $\phi^a$ arises at $O(\eps^2)$, he was able to determine the $O(\eps)$ anomalous dimensions of the operators $\Phi_2$ and $\Phi_2^{ab}$ appearing in the $\phi^a \times \phi^b$ OPE. We are not aware of any further work using Polyakov's approach. It would be very interesting to understand it better and to explore its full potential.

\section{Open problems}
\label{sec:open}

We will conclude by listing several related open problems, in the order of increasing difficulty.

1. Our results should admit a generalization to the fixed point of the $\phi^3$ theory in $6-\eps$ dimensions.\footnote{{\bf Note added:} After this paper was submitted, Yu Nakayama informed us about his unpublished results in this direction.}

2. One should be able to extend our results to the theories with fermions, such as the UV fixed point of the Gross-Neveu model in $2+\eps$ dimensions and the fixed point of the Yukawa model in $4-\eps$ dimensions. These models are described by the Lagrangians
\begin{gather}
\mathcal{L}_{\rm GN}=\bar \psi\! \not\! \del\psi+G(\bar \psi \psi)^2\,,\\
\mathcal{L}_{\rm Y}=\bar \psi\! \not\!\del\psi+\half(\del\phi)^2+ y \phi (\bar \psi \psi) + g \phi^4\,.
\end{gather}
The equations of motion imply that the multiplet of $\psi$, short in the free theory, should recombine with that of $\psi(\bar\psi \psi)$ in Gross-Neveu, and with that of $\psi\phi$ in Yukawa. It appears likely that this recombination can be used to constrain the leading anomalous dimensions. One should consider OPEs which, in the free theory limit, contain the field $\psi$.

3. Let's come back to the WF fixed point, and discuss the recombinations for the higher spin currents, which we already mentioned in footnote \ref{note:spinrec}. The story goes as follows \cite{PedroO(N)}. The free scalar theory contains conserved currents of any even spin of the form\footnote{In the free $O(N)$ model for $N>1$ there are also odd spin conserved currents, of the same form as \reef{eq:Jl} where the two $\phi$'s carry antisymmetrized $O(N)$ indices.}
\beq
J^{(l)}=\phi \overleftrightarrow{\del}_{\!\!\!\mu_1}\cdots \overleftrightarrow{\del}_{\!\!\!\mu_l} \phi-\text{traces}\,.
\label{eq:Jl}
\eeq
All these operators are primaries in the free theory.
The $l=2$ case corresponds to the stress tensor, which remains conserved also at the WF fixed point, and so keeps its canonical dimension $d$. It is a well-known fact that all of the $l>2$ currents acquire anomalous dimensions at the second order in $\eps$ \cite{Wilson:1973jj}:
\beq
\gamma^{(l)}=\Bigr(1-\frac 6{l(l+1)}\Bigl)\frac{\eps^2}{54}+\ldots
\label{eq:gl}
\eeq
 As a consequence, they are not conserved at the WF fixed point.\footnote{This is in agreement with the recent Coleman-Mandula-like theorem \cite{Maldacena:2011jn,Alba:2013yda}, which forbids the existence of conserved higher-spin currents in interacting higher-dimensional CFTs. The theorem has been proven in $d=3,4$, but presumably holds for any dimensions $d>2$, integer or not.} Since current conservation is a shortening condition for its conformal multiplet, we conclude that the multiplets $\{J^{(l)}\}$, short in the free theory, should become long in the interacting theory. They can do this only by eating a multiplet of another primary, which must have an appropriate dimension and spin to match the quantum numbers of $\del\cdot J^{(l)}$.
Thus, in analogy to Eq.~\reef{eq:rec}, we must have
\beq
\{J^{(l)}\}_{\rm WF} \approx \{J^{(l)}\}_{\rm free} + \{\tilde J^{(l-1)}\}_{\rm free}\,,
\label{eq:recl}
\eeq
where $\tilde J^{(l-1)}$ must be a primary operator of spin $(l-1)$ and dimension 
\beq
\Delta_{\tilde J}=\Delta_J+1=l+3\,.
\eeq
Interestingly, one can show that for $l=2$, the free theory contains no primaries of spin $1$ and dimension $5$, and thus recombination \reef{eq:recl} is impossible \cite{PedroO(N)}, \cite{MathProb,MathNb}. In other words, the stress tensor multiplet must remain short, i.e.~conserved, also in the interacting theory. This provides a CFT proof why the WF fixed point must necessarily have an exactly conserved stress tensor operator. On the other hand, for higher spins, one does find candidates for the operator $\tilde J$ with the right quantum numbers.\footnote{This computation was carried out in \cite{PedroO(N)},\cite{MathProb,MathNb} but only for a few low spins, and it would be interesting to prove this for all $l$.}

To summarize, we know from perturbation theory and from the Coleman-Mandula-like theorem \cite{Maldacena:2011jn,Alba:2013yda} that the recombination must occur for $l>2$.  We also know that the free theory contains primaries with the right quantum numbers for this to happen. In analogy with section \ref{sec:computation}, one should be able to use this recombination to reproduce the $O(\eps^2)$ anomalous dimensions of the spin $l$ currents, Eq.~\reef{eq:gl}. This is an open problem worth attacking. The challenge is to find OPEs sensitive to this recombination. For example, the OPE $\phi\times\phi$ does not seem useful, as in the free theory it contains $J^{(l)}$ but not $\tilde J$. 

4. Finally, it would be extremely interesting and important to find a CFT way to compute higher order terms in the $\eps$-expansion. We don't have any concrete proposal for how this can be achieved.
Analysis of three point functions will certainly not suffice for this task, and one will have to consider the full-fledged bootstrap at the four point function level. 

Clearly, the OPE coefficients will also get corrections when one moves to $4-\eps$ dimensions. In this paper we were not sensitive to these corrections; e.g.~in section \ref{sec:computation} all we could say is that $\tilde f/f\to 1$ as $\eps\to 0$. But they will come into play when higher order corrections to the operator dimensions are considered. Notice that corrections to the OPE coefficients are hard to compute from perturbation theory, and almost nothing is known about them.\footnote{The $O(\eps^2)$ correction to the central charge was computed in \cite{Hathrell:1981zb, Jack:1983sk,Cappelli:1990yc,Petkou:1994ad}. See section IV of \cite{El-Showk:2013nia} for a summary.}

\section*{Acknowledgements}
S.R.~thanks Matthijs Hogervorst, Leonardo Rastelli and Balt van Rees for the useful discussions; he especially thanks Sheer El-Showk, Miguel Paulos, and David Simmons-Duffin for the collaboration at the initial stages of this work. Z.M.T. thanks the theory group at CERN for hospitality during the work on this project. 
This research was partly supported by the National Centre of Competence in Research SwissMAP funded by the Swiss National Science Foundation, and also by the \'Ecole normale sup\'erieure in Paris, France, under LabEx ENS-ICFP: ANR-10-LABX-0010/ANR-10-IDEX-0001-02 PSL. Z.M.T also acknowledges support from the Public Service Commission, Singapore, under an Overseas Merit Scholarship.

\appendix

\section{Conformal OPE}
\label{sec:confope}

In this appendix we will study the structure of the conformal OPE \reef{eq:OPEWF}. 
We will have to consider two cases (i) and (ii) defined in Eq.~\reef{eq:twocases}. 

Consider first case (i) when all three operators $V_n$, $V_{n+1}$, $V_1$ are primaries. To simplify the notation, let's consider 
three scalar primary operators $\calA$, $\calB$, $\calC$ of dimensions $a,b,c$, so that we are dealing with the OPE:
\begin{align}
\calA(x) \times \calB(0) &\supset f |x|^{c-a-b} [1+ q_1\, x^\mu \del_\mu + q_2\, x^\mu x^\nu \del_\mu \del_\nu + q_3\, x^2 \Box +\ldots]\calC(0)\,
\label{eq:opeABC}\\ 
&\equiv f |x|^{c-a-b} P(x,\del_y)\calC(y)|_{y=0}\,.\nn
\end{align}
The differential operator $P$ incorporates abstractly all terms in the OPE. The overall OPE coefficient $f$ depends on the CFT, but the differential operator $P$ is universal. It depends only on the transformation properties of the operators under the conformal group. 

{An all-order integral representation for the operator $P$ was given in \cite{Ferrara:1974ny}, Eq. (2.1). For our purposes, we need the expansion of $P$ up to the second order in $x$ shown in \reef{eq:opeABC}.
This can be worked out simply as follows.} Conformal symmetry implies the well-known functional form of the three point function \reef{eq:3pt}. On the other hand, the same correlation function can be computed using the OPE as:
\beq
f P(x-y,\del_y)\langle \calC(y)\calC(z)\rangle, \qquad \langle \calC(y)\calC(z)\rangle = |y-z|^{-2c}\,.
\label{eq:3ptOPE}
\eeq 
Expanding \reef{eq:3pt} in the limit $|x-y|\ll |x-z|$ and matching order by order to \reef{eq:3ptOPE} we can find all coefficients in $P$. The  three coefficients of interest to us are:
\begin{align}
q_1&= (c+a-b)/(2c)\,,\nn\\
q_2 &= (c+a-b)(c+a-b+2)/[8c(c+1)]\,,\nn\\
q_3&= - (c+a-b)(c-a+b)/[16 c(c+1)(c-\delta)],\quad \delta=d/2-1\,.
\label{eq:q3prim}
\end{align}
One interesting and crucial for us difference between these coefficients is that $q_1$ and $q_2$ remain finite in the limit $c\to\delta$, while $q_3$ has a first order pole.\footnote{This patterns continues in higher orders: terms with at least one $\Box$ have a first-order pole for $c\to \delta$, while terms without $\Box$ remain finite. This follows from the mentioned all-order representation of $P$ \cite{Ferrara:1974ny}.}
Substituting $a=\Delta_n$, $b=\Delta_{n+1}$, $c=\Delta_1$, and taking the leading term in the $\eps\to 0$ limit, the expression for $q_3$ reduces to \reef{eq:q}.

We now turn to case (ii), when one of the two operators $V_n$, $V_{n+1}$ is the descendant $V_3$. Continuing the abstract treatment,
we will study the OPE:
\begin{align}
\Box\calA(x) \times \calB(0) &= \tilde f |x|^{c-a-b} [1+ \tilde q_1\, x^\mu \del_\mu + \tilde q_2\, x^\mu x^\nu \del_\mu \del_\nu + \tilde q_3\, x^2 \Box +\ldots]\calC(0)\,.
\end{align}
In practice we will have $\calA=\calC=V_1$ but let's keep it general for the moment. The relative OPE coefficients $\tilde q_i$ can be obtained by simply applying $\Box_x$ to both sides of \reef{eq:opeABC}. We obtain:
\begin{align}
&\tilde q_1=\frac{a+b-c-d}{a+b-c-d+2}q_1\,,\\
&\tilde q_2=\frac{a+b-c-d-2}{a+b-c-d+2}q_2\,,\\
&\tilde q_3=\frac{(a+b-c-d)(a+b-c-2)q_3+2q_2}{(a+b-c)(a+b-c-d+2)}\,.
\end{align}

We will now use these equations to analyze directly the two sub-cases of case (ii):

(a) $n=2$. We have the OPE $V_2\times V_3\supset V_1$. The order of the operators does not matter for the computation of the singular part of $\tilde q_3$ as long as $\tilde q_1, \tilde q_2$ remain nonsingular. Thus we can apply the above result for
\beq
\Box V_1 \times V_2 \supset V_1.
\eeq
We substitute $a=c=\Delta_1=\delta+\gamma_1$, $b=\Delta_2+\gamma_2$, $d=4-\eps$ into the above formulas. 
We will only need to consider the case when $\gamma_1$ is higher order in $\eps$ than $\gamma_2$. In this case 
we find that $\tilde q_1, \tilde q_2$ are indeed nonsingular, while the leading asymptotics for $\tilde q_3$ takes the form:
\beq
\tilde q_3\approx {(\eps-\gamma_2)}/{(16 \gamma_1)}\,.
\label{eq:n2gen}
\eeq
If we further specialize to $\gamma_2\approx y_{2,1}\eps$, $\gamma_1\approx y_{1,2}\eps^2$, we get
\beq
\tilde q_3\approx ({1-y_{2,1}})/({16y_{1,2}\eps})\,.
\label{eq:n2}
\eeq

(b) $n=3$. We have the OPE $V_3\times V_4\supset V_1$. We can apply directly the above result for
\beq
\Box V_1 \times V_4 \supset V_1.
\eeq
We substitute $a=c=\Delta_1=\delta+\gamma_1$, $b=\Delta_4+\gamma_4$, $\gamma_4\approx y_{4,1}\eps$, $\gamma_1\approx y_{1,2}\eps^2$ into the above formulas. 
We find that $\tilde q_1, \tilde q_2$ are nonsingular, while
\beq
\tilde q_3\approx ({y_{4,1}-1})/({16y_{1,2}\eps})\,.
\label{eq:n3}
\eeq

\footnotesize
\bibliography{rec-biblio.bib}
\bibliographystyle{utphys.bst}
\end{document}